\documentclass[twocolumn,showpacs,preprintnumbers,amssymb,pra]{revtex4} 
\usepackage{graphicx}
\usepackage{dcolumn}
\usepackage{color}
\usepackage{bm}
\begin{document} 
\newcommand{\ket}[1]{|#1\rangle}
\newcommand{\bra}[1]{\langle#1|}
\newcommand{\braket}[2]{\langle#1|#2\rangle}
\newcommand{\kb}[2]{|#1\rangle\langle#2|}
\newcommand{\kbs}[3]{|#1\rangle_{#3}\phantom{i}_{#3}\langle#2|}
\newcommand{\kets}[2]{|#1\rangle_{#2}}
\newcommand{\bras}[2]{\phantom{i}_{#2}\langle#1|}
\newcommand{\af}{\alpha}
\newcommand{\bt}{\beta}
\newcommand{\gm}{\gamma}
\newcommand{\la}{\lambda}
\newcommand{\dt}{\delta}
\newcommand{\s}{\sigma}
\newcommand{\qq}{(\s_{y}\otimes\s_{y})}
\newcommand{\uu}{\rho_{12}\qq\rho_{12}^{*}\qq}
\newcommand{\tr}{\textrm{Tr}} 

\title{Atomic entanglement mediated by a squeezed cavity field}
\author{Biplab Ghosh}
\altaffiliation{Email:biplab@bose.res.in}
\affiliation{S. N. Bose National Centre for Basic Sciences,
Salt Lake, Kolkata 700 098, India}
\author{A. S. Majumdar}
\altaffiliation{Email:archan@bose.res.in}
\affiliation{S. N. Bose National Centre for Basic Sciences,
Salt Lake, Kolkata 700 098, India}
\author{N. Nayak}
\altaffiliation{Email:nayak@bose.res.in}
\affiliation{S. N. Bose National Centre for Basic Sciences,
Salt Lake, Kolkata 700 098, India}
\date{\today}

\vskip 0.5cm                              
\begin{abstract}
We consider the coherent state radiation field inside a micromaser
cavity and study the entanglement mediated by it on a pair of
two level atoms passing though the cavity one after the other. We then
investigate the effects of squeezing of the cavity field on the
atomic entanglement. We compute the entanglement of formation for
the emerging mixed two-atom state and show that squeezing of the cavity
radiation field can increase the atomic entanglement.
\end{abstract}                                                                 
\pacs{03.67.-a, 03.65.Ud}
\maketitle    
                                                                 
\section*{I. Introduction}

The most interesting idea associated with composite quantum systems
is quantum entanglement. A pair of particles is said to be entangled in 
quantum mechanics if its state cannot be expressed as a product of the states 
of its individual constituents. This was first noted by Einstein, Podolsky 
and Rosen in 1935\cite{1}. The preparation and manipulation of these 
entangled states that have nonclassical and nonlocal properties leads to a 
better understanding of basic quantum phenomena. For example, complex 
entangled 
states, such as the Greenberger, Horne, and Zeilinger triplets of 
particles\cite{2} are used for tests of quantum nonlocality\cite{3}. 
In addition to pedagogical aspects, entanglement has become a fundamental 
resource
in quantum information processing\cite{4} and there has been rapid development 
of this subject in recent years\cite{5}.

In recent years entanglement has been widely observed within the framework 
of atom-photon interactions such as in optical and microwave 
cavities\cite{6}. 
An example that could be highlighted is the generation of 
a maximally entangled state 
between two modes in a single cavity using a Rydberg atom coherently 
interacting with each mode in turn\cite{7}. The utility of entangled 
atomic qubits for quantum information processing has prompted several
new methods for their generation\cite{8}. In many of these schemes
the transfer of entanglement between two different Hilbert spaces, i.e., from 
the photons to the atoms\cite{9,10}, is involved. The properties of the 
radiation field involved govern the quantitative nature of
atomic entanglement generated through such transfers.

The squeezed radiation field\cite{18} has wide applications in many different
arenas of quantum optics. The relation between squeezing and entanglement 
in general, is
itself an interesting issue which has been discussed through many approaches in
the literature\cite{squeezent}. Squeezing has been used as a resource 
in several
protocols of generating and distilling entanglement, and in information 
transfer\cite{squeezent2}.  In particular, it has been shown 
how atomic
qubits can be entangled with the help of a squeezed radiation field using
one or two optical cavities\cite{squeezent3}. In this Letter we will study 
the effects of squeezing parameters
of a squeezed radiation field inside a microwave cavity on the quantitative 
entanglement of atomic 
qubits passing through it.

The motivation for this work is to investigate the role of squeezing of
the radiation field inside a cavity on the atomic entanglement mediated by it.
We will focus
on a micromaser system\cite{10,21} in which two-level Rydberg atoms are sent 
into the cavity at such a rate that the probability of two atoms being 
present there is negligibly small. Since the atoms do not interact directly 
with each other, the properties
of the radiation field encountered by them bears crucially on the nature
of atomic entanglement. We take the initial state of the two atoms as 
separate or product state, and the emergent two-qubit state is of a mixed
entangled type\cite{10}. The interaction between the atom and the field is 
governed by the
Jaynes-Cummings model\cite{13} which is experimentally realizable.
We consider the cavity to be of a non-leaky
type, i.e., 
$Q=\infty$, and the cavity-QED experiments are very close to such  
situations\cite{6}. We quantify the two-atom entanglement by computing
the entanglement of formation\cite{16}  and demonstrate how the entanglement 
can be increased
by the squeezing of the radiation field, if the average cavity
photon number is kept fixed.

We begin with a brief description of the basic framework. We consider a
single mode cavity and two two-level atoms initially prepared in their
upper excited states $|e_1>$ and $|e_2>$ which pass through the cavity
one after the other. We first consider a coherent state field 
inside the cavity.
A coherent states contains an indefinite number of photons and is a 
minimum uncertainty state\cite{20} standing at the 
threshold of the classical-quantum limit. These states are parametrised by a 
single complex number $\alpha$ as follows:  
\begin{eqnarray}
\ket{\alpha}=\sum_n\frac{\alpha^n}{\sqrt{n!}}\ket{n}.
\label{29}
\end{eqnarray}
A coherent state is an eigenstate of the annihilation operator $a$ written as 
\begin{eqnarray}
a\ket{\alpha}=\alpha\ket{\alpha}
\label{30}
\end{eqnarray}
and obeys a Poissonian distribution function in the photon number 
representation 
given by 
\begin{eqnarray}
P_n=\frac{e^{-<n>}<n>^n}{n!}
\label{31}
\end{eqnarray}
with the average photon number $<n>=|\alpha|^2$. 
The distribution function $P_n$ peaks at non-zero photon number, i.e., 
$n_{peak}\ne0$.

The Janynes-Cummings interaction\cite{13} leads to
a tripartite joint state of the cavity field and the two atoms passing
through it given by
\begin{eqnarray}
\ket{\Psi(t)}_{a-a-f}=\sum_nA_n[\cos^2{(\sqrt{n+1}gt)}
\ket{e_1,e_2,n}\nonumber\\
+\cos{(\sqrt{n+1}gt)}\sin{(\sqrt{n+1}gt)}\ket{e_1,g_2,n+1}\nonumber\\
+\cos{(\sqrt{n+2}gt)}\sin{(\sqrt{n+1}gt)}\ket{g_1,e_2,n+1}\nonumber\\
+\sin{(\sqrt{n+1}gt)}\sin{(\sqrt{n+2}gt)}\ket{g_1,g_2,n+2}]
\label{32}
\end{eqnarray} 
where $P_n=|A_n|^2$ is the photon distribution function of the coherent 
state field.
Since we are interested in calculating the entanglement
of the joint two-atom state after the atoms emerge from the cavity, we 
consider the reduced density matrix $\rho(t)$ of the two-atom state given by
\begin{eqnarray} 
\rho(t)={\textrm Tr}_{\mathrm{field}}({|\Psi(t)>}_{a-a-f.a-a-f}{<\Psi(t)|})
\label{12}
\end{eqnarray}
obtained after taking trace over the field variables.
This state can be written in the matrix form in the basis
of $|e_1>, |e_2>, |g_1>$ and $|g_2>$ states as
\begin{eqnarray}
\rho_{a-a}=\left(\begin{matrix}{\gamma_1&\gamma_7&\gamma_8&\gamma_6 \cr 
\gamma_7&\gamma_2&\gamma_4&\gamma_9\cr \gamma_8&\gamma_4&\gamma_3&\gamma_{10} 
\cr\gamma_6&\gamma_9&\gamma_{10}&\gamma_5}\end{matrix}
\right).
\label{33}
\end{eqnarray}
where
\begin{eqnarray}
\gamma_1&=&\sum_nP_n\cos^4{(\sqrt{n+1}gt)},\nonumber\\
\gamma_2&=&\sum_nP_n\cos^2{(\sqrt{n+1}gt)}\times\nonumber\\
&&\sin^2{(\sqrt{n+1}gt)},\nonumber\\
\gamma_3&=&\sum_nP_n\cos^2{(\sqrt{n+2}gt)}\times\nonumber\\
&&\sin^2{(\sqrt{n+1}gt)},\nonumber\\
\gamma_4&=&\sum_nP_n\sin^2{(\sqrt{n+1}gt)}\times\nonumber\\
&&\cos{(\sqrt{n+1}gt)}\cos{(\sqrt{n+2}gt)},\nonumber\\
\gamma_5&=&\sum_nP_n\sin^2{(\sqrt{n+1}gt)}\times\nonumber\\
&&\sin^2{(\sqrt{n+2}gt)},\nonumber\\
\gamma_6&=&\sum_n\sqrt{P_nP_{n-2}}\cos^2{(\sqrt{n+1}gt)}\times\nonumber\\
&&\sin{(\sqrt{n}gt)}\sin{(\sqrt{n-1}gt)},\nonumber\\
\gamma_7&=&\sum_n\sqrt{P_nP_{n-1}}\cos^2{(\sqrt{n+1}gt)}\times\nonumber\\
&&\cos{(\sqrt{n}gt)}\sin{(\sqrt{n}gt)},\nonumber\\
\gamma_8&=&\sum_n\sqrt{P_nP_{n-1}}\cos^3{(\sqrt{n+1}gt)}\times\nonumber\\
&&\sin{(\sqrt{n}gt)},\nonumber\\
\gamma_9&=&\sum_n\sqrt{P_nP_{n-1}}\sin^2{(\sqrt{n+1}gt)}\times\nonumber\\
&&\cos{(\sqrt{n+1}gt)}\sin{(\sqrt{n}gt)},\nonumber\\
\gamma_{10}&=&\sum_n\sqrt{P_nP_{n-1}}\sin^2{(\sqrt{n+1}gt)}\times\nonumber\\
&&\cos{(\sqrt{n+2}gt)}\sin{(\sqrt{n}gt)}.
\label{34}
\end{eqnarray}

\vskip 1cm

\begin{figure}[h!]
\begin{center}
\includegraphics[width=8cm]{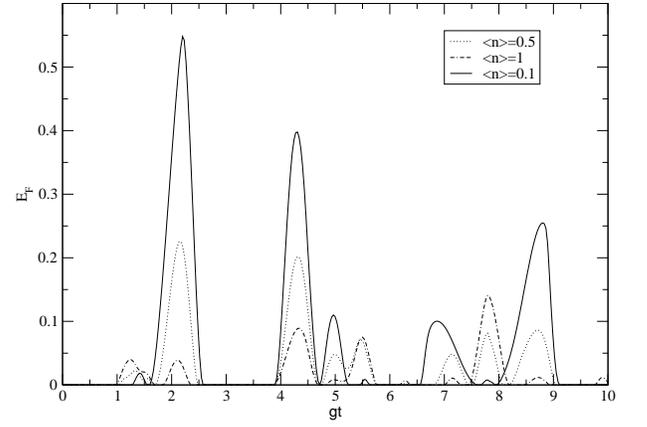}
\caption{Two-atom entanglement mediated by the coherent state cavity field at
low average photon number is plotted versus $gt$.}
\end{center}
\end{figure}

We compute the entanglement of formation $E_F$ for the state
$\rho_{a-a}$, using the Hill-Wootters formula\cite{16}
\begin{eqnarray}
E_{F}(\rho)=h\left(\frac{1+\sqrt{1-C^{2}(\rho)}}{2}\right),
\label{13}
\end{eqnarray}
where $C$ is called the concurrence defined as
\begin{eqnarray}
C(\rho)=\max(0, \sqrt\la_1-\sqrt\la_2-\sqrt\la_3-\sqrt\la_4),
\label{14}
\end{eqnarray}
where the 
$\la_{i}$ are the  eigenvalues of $\uu$ in descending order,
 and 
\begin{eqnarray}
h(x)=-x\log_{2}x-(1-x)\log_{2}(1-x)
\label{15}
\end{eqnarray}    
is the binary entropy function.

\vskip 1cm

\begin{figure}[h!]
\begin{center}
\includegraphics[width=8cm]{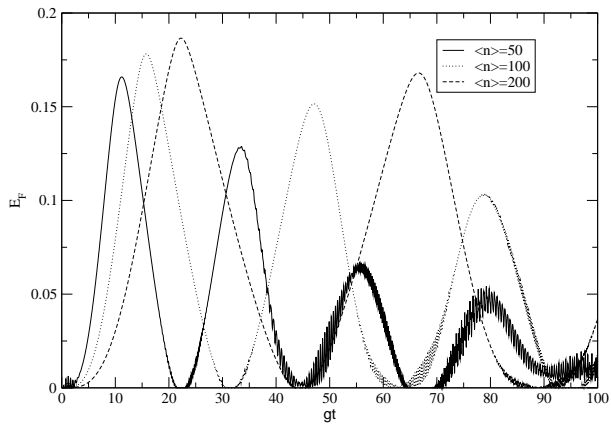}
\caption{Atom-atom entanglement mediated by coherent state cavity field at 
high average photon number is plotted 
versus $gt$.}
\end{center}
\end{figure}

The entanglement of formation $E_F$ is computed separately for low and 
high photon numbers
as the two cases have distinctive features for the coherent state
field. $E_f$ is plotted versus the Rabi angle $gt$ ($g$ is the atom-photon
coupling constant of the Jaynes-Cummings interaction, and $t$ is the time
spent by the atom inside the cavity)
for low average photon number $<n>$ in Figure~1. 
The peaks of the entanglement of formation are reflective of the photon
statistics that are typical in micromaser dynamics\cite{21}.
We see that $E_F$ falls off sharply as $n$ increases. For small photon
numbers, $n_{peak}\approx0$ and hence, the evolution of $E_F$ is similar to 
the case when a thermal field is inside the cavity\cite{ghosh}. For large 
$<n>$, $n_{peak}$ moves 
significantly to the right (Figure~2) and 
its influence is completely different compared to that for the low $<n>$ case.
Quantum effects which are
predominant primarily when the photon number is low, help to increase
the peak value of $E_f$.
We note in Figure~2 that in general, $E_F$ increases slightly with $<n>$ with 
its time evolution being different for different $<n>$.
This is reflective of the collapse-revival characteristic in the 
Jaynes-Cummings model\cite{21}. We further note that though $E_F$ is higher
for the low photon number category (Figure~1), this behaviour is reversed
for the high photon 
number category (Figure~2). For high $<n>$, the features of generated 
entanglement are thus significantly different from those in the 
case of the thermal field\cite{ghosh}.

The above analysis sets the stage for the consideration of a squeezed
radiation field inside the cavity.
A class of minimum-uncertainty states are known as squeezed states.
In general, a squeezed states have less noise in one quadrature than a 
coherent state. To satisfy the requirements of a minimum-uncertainty state 
the noise in the other quadrature is greater than that of a coherent state. 
Coherent states are a particular category of a more general class of 
minimum uncertainty states with equal noise in both quadratures.
Our purpose
here is to study what effect squeezing of the radiation field has on
the entanglement of a pair of atoms passing through it.
The single mode field inside the cavity can be written as
\begin{eqnarray}
E(t)=a_1 \cos{\omega t} + a_2 \sin{\omega t}
\label{35}
\end{eqnarray}
where $a_1=(a+a^\dagger)/2$ and $a_2=(a-a^\dagger)/2i$ are the two 
quadratures satisfying $[a_1,a_2]=i/2$. The variances 
$\Delta a_1=\sqrt{<a_1^2>-<a_1>^2}$ and $\Delta a_2=\sqrt{<a_2^2>-<a_2>^2}$ 
satisfy 
\begin{eqnarray}
\Delta a_1 \Delta a_2 \ge\frac{1}{4}.
\label{36}
\end{eqnarray}
The coherent state or the minimum uncertainty state given by 
Eqs.(\ref{29}-\ref{31}) 
satisfy the equality sign along with
\begin{eqnarray} 
\Delta a_1=\Delta a_2=\frac{1}{2}.
\label{37} 
\end{eqnarray}
Further, either of $\Delta a_1$ or $\Delta a_2$ can be reduced below 
$\frac{1}{2}$ at the expense of the other such that Eq.(\ref{36}) is satisfied,
and radiation fields having such properties are called squeezed fields. 

\vskip 0.6cm

\begin{figure}[h!]
\begin{center}
\includegraphics[width=8cm]{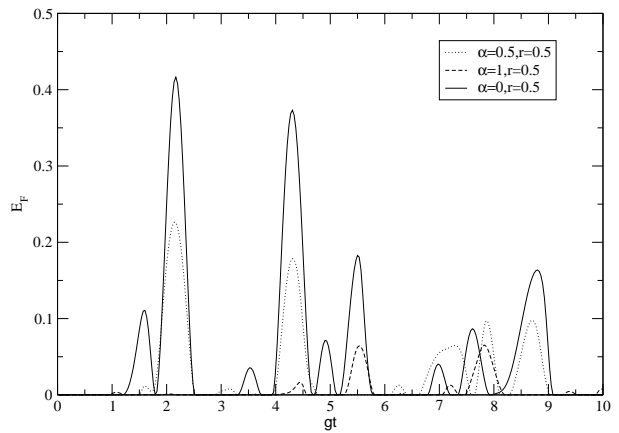}
\caption{Atom-atom entanglement of formation mediated by the squeezed field 
for  different values of $\alpha$ is plotted versus $gt$ for the low
photon number case.}
\end{center}
\end{figure}

The photon distribution function of the squeezed radiation field can be 
represented as  
\begin{eqnarray}
P_n=\frac{1}{n!\mu}(\frac{\nu}{2\mu})^ne^{-\beta^2(\frac{\nu}{\mu}-1)}
|H_n(\frac{\beta}{\sqrt{2\mu\nu}})|^2, 
\label{38}
\end{eqnarray}
where $\beta$ is related to the coherent state amplitude $\alpha$ in 
Eq.(\ref{30}) 
by $\beta=(\mu+\nu)\alpha$ for real $\alpha$. $\mu$ and $\nu$ can be 
represented by the squeezing parameter $r$ as $\mu=\cosh{r}$ and 
$\nu=\sinh{r}$.
The average photon number can thus be written as
\begin{eqnarray}
<n>&=&|\alpha|^2+\sinh^2{r}.
\label{39}
\end{eqnarray}
In terms of the squeezing parameter, the variances of such fields are given by 
\begin{eqnarray}
\Delta a_1=\frac{1}{2}e^{-r}, \nonumber\\
\Delta a_2=\frac{1}{2}e^{r}. 
\label{40}
\end{eqnarray}
Clearly, for $r=0$, the statistics reduce to that for a coherent state 
given by Eq.(\ref{31}). 
$r>0$ gives rise to sub-Poissonian statistics, whereas $r<0$ produces a 
super-Poissonian field. 

\vskip 0.7cm

\begin{figure}[h!]
\begin{center}
\includegraphics[width=8cm]{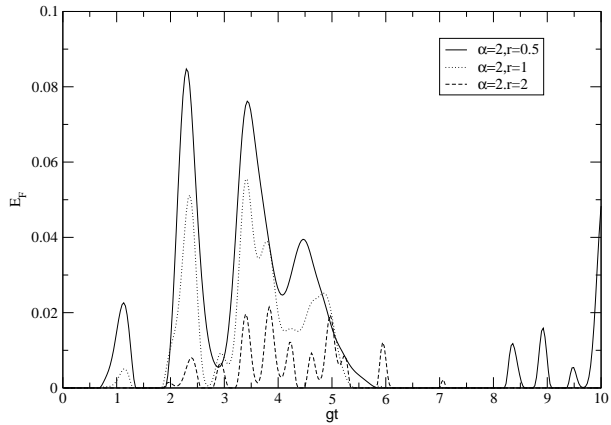}
\caption{$E_F$ mediated by the squeezed field is plotted versus $gt$ for
 different values of the squeezing parameter $r$ corresponding to the
low average photon number case.}
\end{center}
\end{figure} 

As, in the previous case, we first obtain the reduced density matrix
corresponding to the joint two-atom state after passing through a
cavity with the squeezed field. The reduced density matrix has a 
similar form to that of the coherent state field given by Eq.(\ref{33}),
where $\gamma^s$ are also of the same form as given in Eq.(\ref{34}).
The difference in this case arises from the different photon statistics
$P_n$ obtained from the squeezed field distribution function as 
given in Eq.(\ref{38}). 

\vskip 0.6in

\begin{figure}[h!]
\begin{center}
\includegraphics[width=8cm]{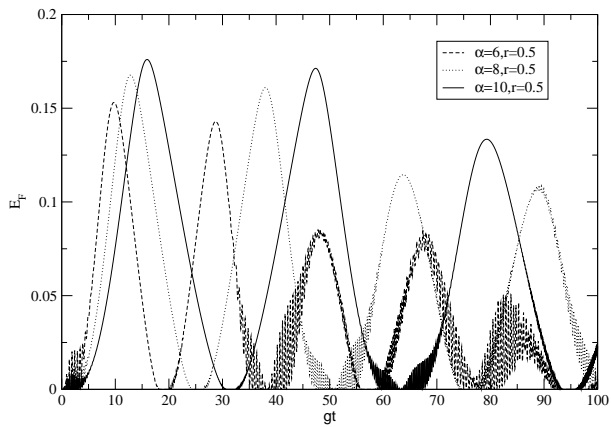}
\caption{$E_F$ mediated by squeezed field for different values of $\alpha$) 
is plotted versus $gt$ for the high average photon number case.}
\end{center}
\end{figure}

The effects of the photon statistics of the squeezed 
field on two-atom entanglement for low average photon number
are displayed in the Figures~3 and 4, for varying $\alpha$ and $r$, 
respectively. 
We see that for low photon numbers, the time evolution of $E_F$ is 
similar to that for a coherent field. 
The effect of the squeezing parameter $r$ enters through $<n>$
in Eq.(\ref{39}). An increase in $r$ increases $<n>$ and thus $E_F$ diminishes 
accordingly. It might appear from Figure~4 that squeezing of the 
radiation field is anti-correlated with the generated atomic entanglement,
but what is actually reflected here is the decrease of $E_F$ caused
by the increase of the average photon number $<n>$. We emphasize on this
point since later (Figure~6) we will indeed see that by squeezing the field
but holding $<n>$ fixed, one can increase the atomic entanglement of 
formation.  
The situation for the high photon number case resembles that
for the coherent state field. This is seen in Figure~5 where a 
larger value of $\alpha$ corresponds to a larger $<n>$, and causes $E_F$
to be slightly increased with increasing $n$ or $\alpha$.

\vskip 0.6cm

\begin{figure}[h!]
\begin{center}
\includegraphics[width=8cm]{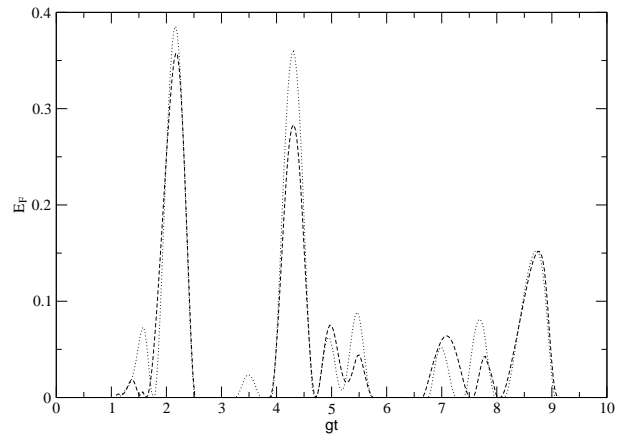}
\caption{Atom-atom entanglement mediated by (i) squeezed cavity field 
(dotted line) when $<n>=0.3$ and $r=0.5$, and  (ii) coherent state field 
(dashed line) when $<n>=0.3$, plotted vs $gt$.}
\end{center}
\end{figure}

The actual effect of squeezing of the cavity field is apparent by performing a
comparitive computation of $E_F$ mediated by the coherent and squeezed 
fields for the same average photon number $<n>$.
In Figures~6 and 7 we plot the two-atom entanglement of formation $E_F$
versus the Rabi angle $gt$ separately for the coherent
state and the squeezed state keeping the average cavity
photon number fixed. In Figure~6 we see that for 
small $<n>$, the dynamics of $E_F$ are similar for both kinds 
of cavity fields. But the striking feature of Figure~6 is in the peaks of $E_F$
for various values of $gt$.  Note that $E_F$ for the squeezed field (dotted
line) is higher compared to the coherent state field (dashed line). Thus
squeezing of the radiation field as represented by the non-vanishing value 
of the 
squeezing parameter $r$, leads to a notable increase in the magnitude
of atomic entanglement over the case the coherent state field 
($r=0$; no squeezing). This trend is also visible in the high photon
number case (Figure~7), though not for all values of $gt$. 

\vskip 0.6cm

\begin{figure}[h!]
\begin{center}
\includegraphics[width=8cm]{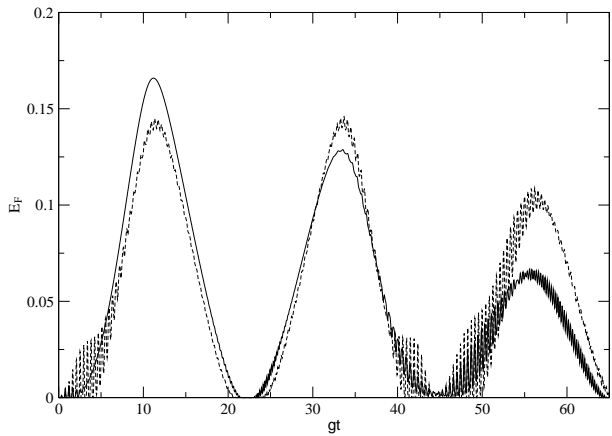}
\caption{Atom-atom entanglement mediated by (i) squeezed cavity field 
(dashed line) when $<n>=50$ and $r=1$, and (ii) coherent state field 
(solid line) when $<n>=50$
when $<n>=50$, plotted vs $gt$.}
\end{center}
\end{figure}

To summarize, in this Letter we have considered a 
micromaser model where two 
spatially separated atoms are entangled via a cavity field. 
The entanglement between the two separate atoms builds up via atom-photon 
interactions inside the cavity, even though no single atom interacts 
directly with another. We have computed the two-atom entanglement as
measured by the entanglement of formation $E_F$ for the cases of the 
coherent state field and the squeezed radiation field inside the cavity. 
Our purpose has
been to perform a quantitative study of the effects of squeezing of the 
bosonic radiation field
on the mediation  of the mixed state entanglement of two atomic qubits.
Two distinct patterns of
entanglement are seen to emerge for the cases corresponding to low and
high average cavity photon numbers, respectively. In the former case the
quantum nature of the radiation field plays a prominent role in enhancing
atomic entanglement with the decrease of $<n>$. The situation reverses
for high $<n>$ case where actually the increase of $<n>$ leads to a slight
increase of $E_F$. 
The key feature prominently observed for the low $<n>$ case
is that the two-atom entanglement can
be increased with squeezing of the cavity field if the average cavity 
photon number is held fixed. Further interesting directions could
be to study the impact of squeezed radiation on the ``monogamous''
character\cite{29} of atomic entanglement, and also to investigate 
the possibility of generating maximally entangled mixed atomic 
qubits\cite{mems}
using squeezing of the bosonic field as a resource.

\end{document}